\documentclass[usegraphicx]{mn2e}
\usepackage{times}
\usepackage{amssymb}
\usepackage{amsmath}

\begin{document}

\title[Long GRBs and Their Hosts at High $z$]
{Long Gamma-Ray Bursts and Their Host Galaxies at High
Redshift}
\author[Lapi et al.]{\parbox[]{6.in}
{A. Lapi$^{1,2}$, N. Kawakatu$^{3,2}$, Z. Bosnjak$^{4,2}$, A.
Celotti$^{2,5}$, A. Bressan$^{6,2,7}$, G.L. Granato$^6$ and L.
Danese$^2$\\ \footnotesize $^1$Dip. Fisica, Univ. `Tor
Vergata', Via Ricerca Scientifica
1, 00133 Roma, Italy\\
$^2$SISSA/ISAS, Via Beirut 2-4, 34014 Trieste, Italy\\
$^3$National Astronomical Observatory of Japan, Mitaka, Tokyo 181-8588, Japan\\
$^4$Institut d' Astrophysique de Paris, bd Arago 98bis,
75014 Paris, France\\
$^5$INFN-Sez. di Trieste, Via Valerio 2, 34100 Trieste, Italy\\
$^6$INAF-Ossevatorio Astronomico di Padova,
Vicolo dell'Osservatorio 5, 35122 Padova, Italy\\
$^7$INAOE, Luis Enrique Erro 1, 72840 Tonantzintla, Puebla,
Mexico}}

\maketitle

\begin{abstract}
Motivated by the recent observational and theoretical evidence
that long Gamma-Ray Bursts (GRBs) are likely associated with
low metallicity, rapidly rotating massive stars, we examine the
cosmological star formation rate (SFR) below a critical
metallicity $Z_{\rm crit}\sim Z_{\odot}/10 - Z_{\odot}/5$, to
estimate the event rate of high-redshift long GRB progenitors.
To this purpose, we exploit a galaxy formation scenario already
successfully tested on a wealth of observational data on
(proto)spheroids, Lyman break galaxies, Ly$\alpha$ emitters,
submm galaxies, quasars, and local early-type galaxies. We find
that the predicted rate of long GRBs amounts to about $300$
events yr$^{-1}$ sr$^{-1}$, of which about $30$ per cent occur
at $z\ga 6$. Correspondingly, the GRB number counts well agree
with the bright {\it SWIFT} data, without the need for an
intrinsic luminosity evolution. Moreover, the above framework
enables us to predict properties of the GRB host galaxies. Most
GRBs are associated with low mass galaxy halos $M_{\rm H}\la
10^{11}\,M_{\odot}$, and effectively trace the formation of
small galaxies in such halos. The hosts are young, with age
smaller than $5\times 10^{7}$ yr, gas rich, but poorly
extincted ($A_V\la 0.1$) because of their chemical immaturity;
this also implies high specific SFR and quite extreme
$\alpha$-enhancement. Only the minority of hosts residing in
large halos with $M_{\rm H}\ga 10^{12}\,M_{\odot}$ have larger
extinction ($A_V\sim 0.7-1$), SFRs exceeding $100\,M_{\odot}$
yr$^{-1}$ and can be detected at submm wavelengths. Most of the
hosts have UV magnitudes in the range $-20\la M_{1350}\la -16$,
and Ly$\alpha$ luminosity in the range $2\times 10^{40}\la
L_{\rm Ly\alpha}\la 2\times 10^{42}$ erg s$^{-1}$. GRB hosts
are thus tracing the faint end of the luminosity function of
Lyman break galaxies and Ly$\alpha$ emitters. Finally, our
results imply that the population of `dark' GRBs occur mostly
in faint hosts at high redshift, rather than in dusty hosts at
low redshift.
\end{abstract}

\section{Introduction}

The spectroscopic detection of the energetic supernova SN
2003dh coincident with GRB 030329 (Hjorth et al. 2003; Stanek
et al. 2003) has firmly established that - at least some - long
GRBs accompany the core collapse of massive stars, as it was
first suggested by the spatial and temporal coincidence of GRB
980425 and SN 1998bw (Galama et al.  1998). Such spectroscopic
signatures of supernovae (SNae) associated with GRBs have been
detected in a handful of cases during the last years, e.g. GRB
031203/SN 2003lw (Malesani et al. 2004), GRB 021211/ SN 2002lt
(Della Valle et al. 2003), GRB 050525A/SN 2005nc (Della Valle
et al. 2006), and the recent case of GRB 060218/SN 2006aj
(Campana et al. 2006, Modjaz et al. 2006). The possibility that
core-collapse SNae are progenitors of GRBs has been further
supported by the detection of re-brightening in the late-time
afterglow light curves, interpreted as a contribution of
accompanying SNae (Bloom et al. 1999, Zeh et al. 2004,
Castro-Tirado \& Gorosabel 1999) and the localizations of
afterglows in star forming regions (e.g. Djorgovski et al.
2001, Bloom et al. 2002, Fynbo et al. 2000, Fruchter et al.
1999, Holland \& Hjorth 1999).

Indeed, the currently mostly favored scenario for the origin of
long GRBs involves the collapse of a massive Wolf-Rayet star
endowed with rotation (Woosley 1993, MacFadyen \& Woosley 1999,
Woosley \& Heger 2004). Recently, Yoon \& Langer (2005)
considered the evolution of massive, magnetized stars where
rapid rotation induces an almost chemically homogeneous
evolution, and found that the requirements of this collapsar
model are satisfied if the metallicity is sufficiently small,
namely less than $0.1\, Z_{\odot}$ (see Woosley \& Heger 2006;
Yoon et al. 2006). This is broadly consistent with the
estimates of metallicities of long GRB hosts, which yielded
preferentially subsolar - down to $10^{-2}\, Z_{\odot}$ -
values (e.g. Gorosabel et al. 2005; Chen et al. 2005; Starling
et al. 2005).

Clearly this scenario has key implications not only on the
physics of the event and on the evolution of massive stars at
low metallicities, but also on the event rate and redshift
distribution of long GRBs compared with that of SNae. The
cosmological consequences of a metallicty threshold can be
explored by considering an average cosmic metallicity
evolution, as in the work by Langer \& Norman (2006). In the
present paper we instead explore the effects of a metallicity
threshold following the star formation and chemical evolution
of individual galaxies.

In order to derive the rates of progenitors and the
characteristics of their host galaxies at high redshift, the
star formation history and the chemical evolution for a large
range of galaxy mass and virialization redshift must be
computed. Baryon condensation in cold gas, stars and in a
central massive black hole (BH) within galaxy Dark Matter (DM)
halos is a quite complex outcome of a number of physical
processes (including shock waves, radiative and shock heating,
viscosity, radiative cooling; star formation, BH accretion, gas
inflow and outflow) largely affecting each other (see Granato
et al. 2001, 2004; Croton et al. 2006; De Lucia et al. 2006).
When treating these processes within a self-consistent
cosmological framework of galaxy formation, most of the
complexity is related to the different scales involved.

For the sake of definiteness, here we adopt as a reference the
galaxy formation scenario developed by Granato et al. (2004),
which consistently accounts for the coevolution of spheroidal
galaxies and their nuclear activity and intrinsically follows
in time the gas content, star formation rate (SFR) and
metallicity evolution for each galaxy mass. This enables us to
investigate in detail the effect of a metallicity threshold on
the properties not only of the GRB population, but also of
their host galaxies.

The outline of this paper is the following. In \S~2 we briefly
review the conceptual issues of the adopted galaxy formation
scenario and explain how the GRB progenitor rate has been
estimated. Our results are presented in \S~3. In \S~4 we
discuss our findings, by comparing them with observational
results and with previous studies. In \S~5 we summarize our
conclusions. Throughout the paper a flat cosmological model
with matter density parameter $\Omega_M=0.27$ and Hubble
constant $H_{0}=72$ km s$^{-1}$ Mpc$^{-1}$ is adopted.

\section{Modelling}

\subsection{Overview of the galaxy formation scenario}

Long GRBs at high redshift ($z\ga 1$) have progenitors which
formed at least $8$ Gyr ago. Their coeval stellar populations
are as old as the populations of spheroidal galaxies and spiral
bulges and older than the populations in present galaxy discs
(see Renzini 2006): thus high redshift long GRBs trace the
formation history of the oldest stellar populations. Low
redshift GRBs conversely trace the star formation in small mass
irregular/interacting galaxies with low SFR (Fruchter et al.
2006; Wainwright et al. 2007). Since we focus on high-$z$ GRBs,
in the following we will neglect the contribution of star
formation in discs of present-day spiral galaxies.

We exploit the physical model elaborated by Granato et al.
(2001, 2004), which follows the evolution of baryons within
protogalactic spheroids taking into account the effects of the
energy fed back to the intra-galactic gas by SN explosions and
by accretion onto the nuclear supermassive BH (see also De
Lucia et al. 2006; Croton et al. 2006). The model envisages
that during or soon after the formation of the host DM halo,
the baryons falling into the newly created potential well are
shock-heated to the virial temperature. The hot gas is
(moderately) clumpy and cools fast especially in the denser
central regions, yielding a strong burst of star formation.
Star formation also promotes the storage of the cooled gas into
a reservoir around the central seed BH, eventually leading to
accretion onto it (see Kawakatu \& Umemura 2002).  The ensuing
SN explosions and the nuclear activity feed energy back to the
baryons, and regulate the ongoing SFR and BH growth. These
mutual energy feedbacks actually \emph{reverse} the formation
sequence of the baryonic component of galaxies compared to that
of DM halos: the star formation and the buildup of central BHs
are completed more rapidly in the more massive haloes, thus
accounting for the phenomenon now commonly referred to as
\emph{downsizing} (e.g. Cowie et al. 1996; Kodama et al. 2004;
Glazebrook et al. 2004).

In Appendix A we present a simplified version of basic
equations of the model, which allow to derive analytical
solutions for the time evolution of SFR, mass in stars and
metallicity, the quantities relevant to this work. These
analytical functions are very good approximations of the more
complex system of equations numerically solved in Granato et
al. (2004; see for details Lapi et al. 2006 and Mao et al.
2007). Because of their fundamental character, these equations
catch the basic aspects of the physical processes ruling star
formation in protogalaxies at high redshift. The adopted
Initial Mass Function (IMF) is a double power-law with slope
$1.25$ from $120\, M_{\odot}$ to $1\, M_{\odot}$ and $0.4$ from
$1\, M_{\odot}$ down to $0.1\, M_{\odot}$ (Romano et al. 2002),
which is quite similar to that proposed by Chabrier (2005).

In Fig.~1 we present the results for DM halos virialized at
$z=6$ and endowed with mass ranging from $10^{10}$ to
$10^{13}\, M_{\odot}$. For masses $M_{\rm H}\ga 10^{12}\,
M_{\odot}$ the SFR increases almost linearly with galaxy age in
the initial stages, and then it is suddenly halted by the
energy feedback from the quasar after a few $10^8$ yr. On the
contrary, for $M_{\rm H}\la 10^{12}\, M_{\odot}$ it is first
almost constant and then slowly declines due to gas exhaustion.
In fact, for small masses the effect of SNae feedback regulates
the star formation, while the BH is rather small there and the
SFR can proceed for a much longer time (see Eq.~[16] in Shankar
et al. 2006).

The mass cycled through stars at any time is easily obtained by
integrating the SFR (see Fig.~1). Note that the quantities
plotted in Fig.~1 refer to redshift $z=6$; however, both SFR
and mass in stars scale approximately as $(1+z)^{3/2}$ for a
given halo mass.

From the figure it is apparent that the chemical enrichment of
the cold gas component is very rapid; e.g. with the adopted IMF
the gas attains $1/100$ and $1/10$ the solar abundance in about
$1.2\times 10^7$ and $5\times 10^7$ yr respectively, almost
independently of the halo mass and redshift. A Salpeter IMF
yields timescales about twice as long. This rapid enrichment is
due to the fast evolution and formation timescales ($t_\star$,
see Appendix A) for the massive stars (greater than $10
\,M_\odot$) relative to the timescale for the infall of the
diffuse medium with primordial composition, which dilutes the
metallicity of the cold star-forming gas. This behavior shows
that a possible change of the IMF at metallicity lower than a
threshold around $3-5\times 10^{-4}\, Z_{\odot}$ (see Bromm et
al. 2001; Schneider et al. 2006) is not critical for the issue
related to long GRBs. In other words, if long GRBs are
associated with low metallicity environments, the quick
enrichment implies that the most relevant epoch for GRB
progenitors comes soon after halo virialization.

This galaxy formation model neglects spatial resolution and
assumes instantaneous mixing, i.e. it averages both SFR and
chemical composition over the entire mass of cold gas.  We
stress that the cold gas is only a small fraction of the
overall baryons associated with the galaxy halo (cf. Appendix
A), and thus the averaging concerns the mass/volume of the
protogalaxy wherein star formation is occurring and not the
overall DM halo mass/volume. On the other hand, metallicity
gradients have been observed in local spheroidal galaxies,
$\Delta [Z/H]/\Delta \log r \approx -0.25$ (e.g. Annibali et
al. 2007, Sanchez-Blanquez et al. 2007). This implies
variations of about a factor two in metal abundance within a
radius enclosing most of the galaxy mass. As we shall see such
a factor is not crucial to the conclusions of this paper.

\begin{figure}
\begin{center}
\includegraphics[width=6truecm, height=16 truecm]{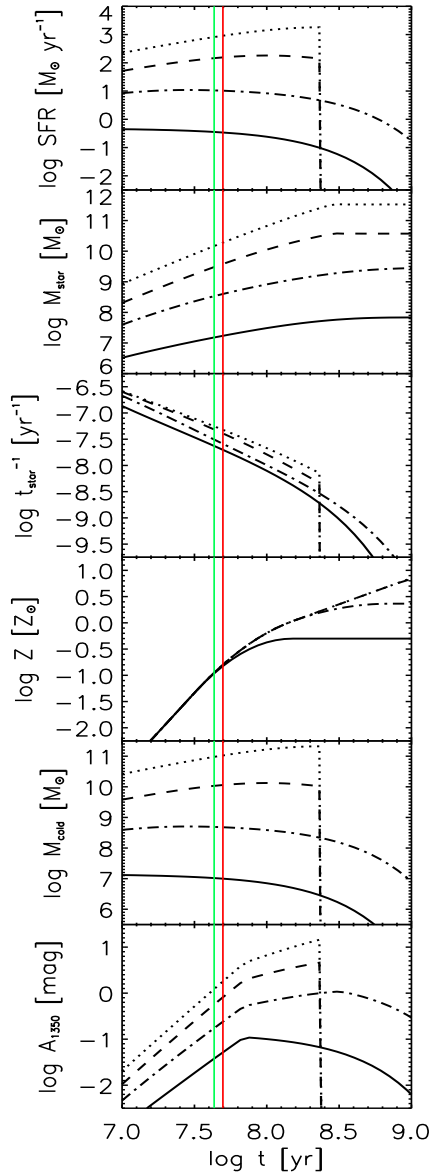}
\caption{SFR, stellar mass, specific SFR,
average cold gas metallicity, cold gas mass, and extinction at $1350$
\AA~ (from top to bottom) as a function of the galactic age for halos
of masses $10^{10}\,M_{\odot}$ ({\it solid} lines), $10^{11}\,M_{\odot}$
({\it dot-dashed} lines), $10^{12}\,M_{\odot}$ ({\it dashed} lines),
$10^{13}\,M_{\odot}$ ({\it dotted} lines), virialized at redshift $z =
6$. The vertical lines mark the epoch when the cold gas metallicity
attains the critical thresholds $Z_{\odot}/10$ ({\it green}) and
$Z_{\odot}/5$ ({\it red}).}
\end{center}
\end{figure}

The model here adopted has the very valuable asset that
successfully fits a wealth of observational data and
constraints regarding protospheroids, Lyman break galaxies and
Ly$\alpha$ emitters, submm-selected galaxies, quasars, EROS and
local early-type galaxies; for a detailed comparison with the
observational data, we defer the interested reader to the
papers by Granato et al. (2001, 2004, 2006), Cirasuolo et al.
(2005), Silva et al. (2005), Lapi et al. (2006, in particular
their Table 2), Mao et al. (2007); furthermore, clustering
properties of submm galaxies have been extensively discussed in
the context of our model by Negrello et al. (2007).

\subsection{Estimate of long GRB progenitor rates}

The key ingredients provided by the galaxy formation model
related to this work are the SFR $\dot{M}_{\star}(t)$ and the
average gas metallicity $Z(t)$ as a function of the age $t$ for
an individual galaxy within a given halo mass $M_{\rm H}$
virialized at redshift $z$ (see Appendix A for handy
approximations). Notice that $Z$ refers to the average
metallicity of the gaseous component from which new stars form.

The cosmic SFR per unit volume at redshift $z$ contributed by
objects with average gas metallicity $Z < Z_{\rm crit}$ is thus
given by
\begin{equation}
\mathcal{SFR}(z)_{Z<Z_{\rm crit}} = \int{\rm d} \dot{M}_{\star}
{\rm d} M_{\rm H}\, \frac{\mathrm{d}^2\, N_{\mathrm{ST}}}{\mathrm{d}
M_{\rm H}\, \mathrm{d} t_z}\, \frac{\mathrm{d}}{\mathrm{d}\ln
\dot{M}_{\star}} \mathcal{T}^{>\dot{M}_{\star}}_{Z<Z_{\rm crit}}~.
\end{equation}
where $\mathrm{d}^2\, N_{\mathrm{ST}}/\mathrm{d} M_{\rm H}\,
\mathrm{d} t_{\rm z}$ are the formation rates of DM halos at
cosmic time $t_{\rm z}$ computed using the Sheth \& Tormen
(1999) mass function, and
$\mathcal{T}^{>\dot{M}_{\star}}_{Z<Z_{\rm crit}}$ is the time
the galaxy spends at SFR higher than $\dot{M}_{\star}$ and
metallicity lower than $Z_{\rm crit}$.

It is plausible that the core-collapse, believed to give rise
to the formation of a BH and a GRB event, is related to the
final phases of the evolution of stars with masses greater than
$12\,M_{\odot}$ and with rapidly rotating cores (Yoon \& Langer
2005; Woosley \& Heger 2006; Yoon et al. 2006). From a
theoretical point of view a major issue in this respect is the
large mass loss, that would entail also large angular momentum
loss. This problem is possibly alleviated in stars of
metallicity below a critical threshold lower than $1/5-1/3\,
Z_{\odot}$ and high initial spin rate (Woosley \& Heger 2006;
Yoon et al. 2006). Though observations are still scanty in
supporting this expectation, we will consider low metallicity
and high rotation velocity as necessary and sufficient
conditions for a massive star to be considered as a GRB
progenitor.

The GRB progenitor rate will thus be based on the above
preliminary theoretical estimates which suggest that the
expected fraction of GRB progenitors with respect to the number
of massive stars ($8\, M_{\odot}\la m_\star\la 100\,M_{\odot}$)
is $f_{\rm prog}\ga 2-3$ per cent for metallicity $Z_{\rm crit}
\la Z_{\odot}/5$ (see Woosley \& Heger 2006; Yoon et al. 2006;
see also Bissaldi et al. 2007).

The absolute GRB progenitors rate per unit volume can be then
expressed as
\begin{equation}
\mathcal{R}_{\rm prog}(z)= f_{\rm prog}\, n_{\rm SN} \,
\mathcal{SFR}(z)_{Z< Z_{\rm crit}},
\end{equation}
where $n_{\rm SN}\equiv\int_{8}^{100}\phi(m_{\star}) {\rm d}
m_{\star} /\int^{100}_{0.1}\, m_{\star}\, \phi(m_{\star}) \,
\mathrm{d} m_{\star}$ is the number of massive stars ending in
SNae per unit mass of formed stars; for the adopted IMF $n_{\rm
SN}\approx 1.4\times 10^{-2}\,M_{\odot}^{-1}$ (it halves for a
Salpeter IMF). Thus assuming $f_{\rm prog}\approx 0.02$, the
number of GRB progenitors per unit mass of formed stars amounts
to about $3.6 \times 10^{-4}\,M_{\odot}^{-1}$ for $Z_{\rm
crit}\la Z_{\odot}/5$.

\subsection{Long GRB number counts}

The number of GRBs at redshift greater than $z$ is
\begin{equation}
\mathcal{R}^{\mathrm{obs}}_{\rm GRB}(>z)\approx \, f_{\rm
beam}\,\int_z{\mathrm{d}z'}~\frac{\mathcal{R}_{\mathrm{prog}}(z')}{1+z'}\,
\frac{\mathrm{d}V}{\mathrm{d}z'}~,
\end{equation}
where $V$ is the cosmological volume per unit solid angle and
the factor $(1+z)^{-1}$ accounts for time dilation effects due
to redshift. Moreover, recall that GRBs are believed to be
anisotropic phenomena owed to the flow collimation and/or
relativistic beaming effects. Thus only a fraction $f_{\rm
beam}$ of the estimated progenitors prompt emission would point
toward us within an opening angle $2\,\theta$ and could then be
observed as a GRB. For the events for which a (jet) opening
angle have been estimated\footnote{These estimates maybe
clearly affected by selection effects, though not easy to
quantify.} assuming that an afterglow light curve break owed to
a jetted structure with a half angle $\theta$ (Ghirlanda et al.
2007), the median value $\theta\approx 6$ deg has been inferred
(see also Guetta et al 2005); correspondingly,  $f_{\rm
beam}\approx 5.5\times 10^{-3}\, (\theta/6\,\mathrm{deg})^2$.
The expected number of long GRBs per unit mass of formed stars
within metal poor environments ($Z\la Z_{\rm crit}$) is thus
$k\approx 1.5\times 10^{-6}\, (n_{\rm SN}/1.4 \times 10^{-2})\,
(f_{\rm prog}/0.02)\,(\theta/6\,\mathrm{deg})^2\,
M_{\odot}^{-1}$.

The long GRB number counts at flux limit $S_{\rm lim}$ and
redshift $> z$ is given as
\begin{equation}
\mathcal{R}^{\mathrm{obs}}_{\rm
GRB}(>z)_{S>S_{\rm lim}}\approx f_{\rm beam}
\int_z{\mathrm{d}z'}~\frac{\mathcal{R}_{\mathrm{prog}}(z')}{1+z'}\,
\frac{\mathrm{d}V}{\mathrm{d}z'}\int_{L_{\rm lim}}^{\infty}
P(L)dL~;
\end{equation}
in this expression $P(L)$ is the equivalent isotropic
luminosity distribution (see \S~3.2 for details), and $L_{\rm
lim}$ is the luminosity corresponding to the limiting flux,
given by
\begin{equation}
L_{\rm lim}(z)=\frac{4\pi\,d^2_L(z)}{K(z)}\, S_{\rm lim},
\end{equation}
in terms of the luminosity distance $d_L(z)$ and of the
$K$-correction $K(z)$; note that the latter quantity depends on
the GRB spectra, see \S~3.2 for details.

\section{Results}

\subsection{Long GRB progenitor rates and redshift distribution}

If progenitors of long GRBs are metal poor massive stars, the
metal abundance $Z$ of the cold gas wherein stars form is a
crucial physical parameter. Fig.~2 shows the cosmological SFR
for the overall galaxy population and for systems with $Z\la
Z_{\rm crit}\approx 0.1$ and $0.2\, Z_{\odot}$ as function of
redshift. The model prediction can be fitted with the
approximate formula
\begin{equation}
\log{\left({\mathcal{SFR}(z)_{Z<Z_{\rm crit}}\over M_{\odot}\, {\rm yr}^{-1}\, {\rm Mpc}^{-3}}\right)}=a+b\,(z-z_{\rm max})^2~,
\end{equation}
where $a=-1.4\, (-3.2)$, $b=-3.8\, (-3.2)\times 10^{-2}$ and
$z_{\rm max}= 4.5\, (6.5)$ for $Z_{\rm crit}=0\,
(Z_{\odot}/10)$.

The predicted overall SFR reproduces fairly well the
extinction-corrected data at $2\la z \la 6$, with reasonable
correction for dust extinction. At $z \la 1$ the observed SFR
is underestimated due to the fact that intrinsically our model
does not account for the contribution of star formation in
discs of present spiral galaxies. On the other hand, only a
minor fraction (less than $20$ per cent) of long GRBs with
redshift determination (though it is difficult to quantify the
observational biases which certainly affect this percentage)
are located at $z\la $1, while we are interested in the bulk of
the (high redshift) GRB population (cf. \S~2.1).

\begin{figure}
\includegraphics[width=\columnwidth]{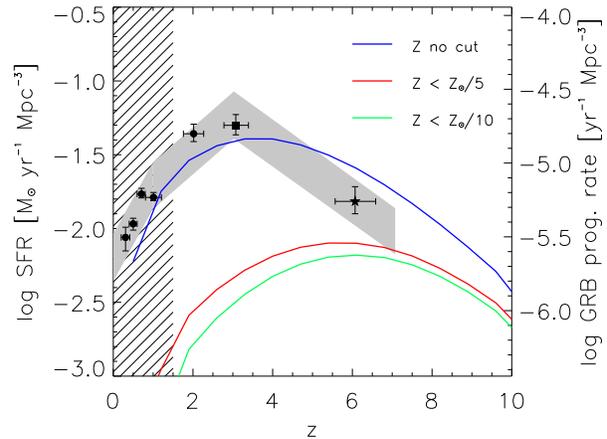}
\caption{Cosmological SFR (left y-axis) and rate of long GRB
progenitors (right y-axis) as a function of redshift, computed with no
threshold on metallicity (\textit{blue} line), and with
thresholds at $Z_{\odot}/5$ (\textit{red} line) and
$Z_{\odot}/10$ (\textit{green} line). Data are from Schiminovich
et al. (2005; \textit{circles}), Steidel et al. (1999;
\textit{squares}), and Bouwens et al. (2006; \textit{stars}). The
\textit{shaded} area illustrates the uncertainties due to the
extinction correction; the data have been rescaled down by a factor of
about $2$ to account for the adopted IMF (see text). In the
\textit{hatched} region the SFR is dominated by discs of present
spiral galaxies, not treated in our model.}
\end{figure}

As expected the SFR in GRB hosts decreases with $Z_{\rm crit}$,
but the effect is differential with redshift. Since the
timescale $t_{\rm crit}\la 5\times 10^7$ yr to reach the
critical metallicity threshold is almost independent of mass
and redshift (cf. Fig.~1), its ratio to the cosmic time
increases toward high $z$ for all halos. The fraction of
overall cosmic star formation occurring in metal poor
protogalaxies raises with redshift, and so does the predicted
rate of GRB progenitors. The net effect is that the expected
redshift distribution of long GRB progenitors peaks to a
redshift significantly higher than that of the cosmic SFR.  As
a consequence the higher the redshift, the more directly GRBs
mirror the cosmic SFR. The fall off of star formation beyond
$z\ga 10$ in the galaxy formation model is a consequence of the
effect of the SN feedback; at increasing $z$ the decrease of
star formation $\propto M_{\rm H}^{1.5}$ is not balanced by the
rise in the number of virialized halos.

\begin{figure}
\includegraphics[width=\columnwidth]{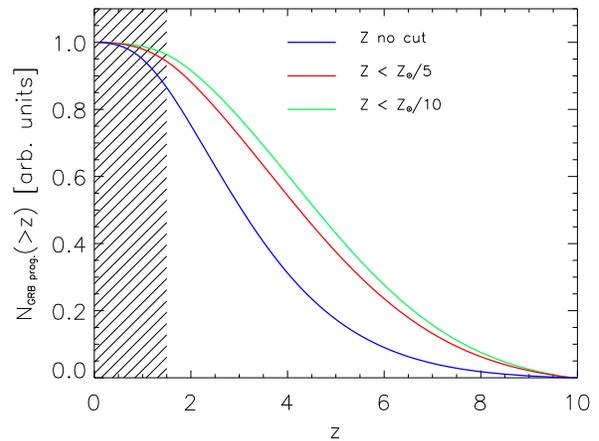}
\caption{Normalized redshift distribution of GRB progenitors, computed
with no threshold on metallicity (\textit{blue} line), and with
thresholds at $Z_{\odot}/5$ (\textit{red} line) and
$Z_{\odot}/10$ (\textit{green} line).}
\end{figure}

\begin{figure}
\includegraphics[width=\columnwidth]{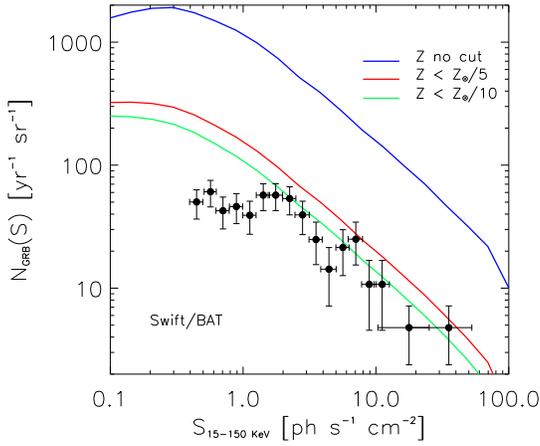}
\caption{GRB counts in the $15-150$ keV band, computed with no
threshold on metallicity (\textit{blue} line), and with
thresholds at $Z_{\odot}/5$ (\textit{red} line) and
$Z_{\odot}/10$ (\textit{green} line). Data (\textit{filled
dots}) are from the $2$-yr {\it SWIFT} catalog.}
\end{figure}

The long GRB progenitor rate shown in Fig.~2 can be integrated
over cosmic time to obtain the normalized redshift distribution
reproduced Fig.~3. For the case $Z_{\rm crit}=Z_{\odot}/10$,
the fraction of progenitors is around $60$ per cent at $z\ga 4$
and around $30$ per cent at $z\ga 6$.

If the threshold $Z_{\rm crit}$ were to refer to stellar rather
than ISM metallicity, the time within which GRB progenitors can
be produced in any galaxy halo would be overestimated by a
factor around $2$. This would in turn double the duty cycle
and, hence, the number of progenitors. The difference in the
predictions worsens with higher $Z_{\rm crit}$ threshold, to
the point where the threshold might never be reached in small
galaxies, which then can host GRB progenitors over the whole
Hubble time. Again the effect is differential with redshift,
with a larger increase in the progenitor number at higher $z$.

\subsection{Long GRB counts with {\it SWIFT}}

In order to assess whether a scenario in which single metal
poor, rotating massive stars are the GRB progenitors is tenable
within the adopted galaxy model framework, we estimated the
predicted long GRB counts which should have been detected by
{\it SWIFT} (Gehrels et al. 2004).

As discussed in \S~2.2 and \S~2.3 we consider a number of
observed GRBs per unit mass $k\approx 1.9\times 10^{-6}\,
M_{\odot}^{-1}$ and a metallicity threshold $Z_{\rm
crit}\approx 1/10-1/5\, Z_{\odot}$.

In order to estimate the observable number of GRBs as function of
limiting flux, we assume that the prompt GRB luminosity distribution
can be characterized, independently of redshift, as
\begin{equation}
P(L)\propto L^{-\delta}\,\mathrm{e}^{-L_{\rm c}/L}~,
\end{equation}
with a low luminosity cutoff at $L_{\rm c}=3\times 10^{51}$ erg
s$^{-1}$, and slope $\delta=2$. This parametrization is
consistent with those constrained by Daigne et al. (2006) and
Guetta et al. (2005).  As for the spectrum, we adopted - as
commonly done - a typical Band representation, with low and
high energy slopes $\alpha=-1$ and $\beta=-2.25$, respectively.
The peak energy is considered to follow a log-normal
distribution (Preece et al. 2000) with mean $\log E_{\rm
peak,0}$=2.74 and dispersion $\sigma=0.3$ dex (see Daigne et
al. 2006). Although such spectral parameters have been
estimated from bright {\it BATSE} GRBs (Preece et al. 2000),
this provides the simplest hypothesis which - given the
relatively small effect owed to the $K$-correction - appears
adequate for the consistency check we intended to perform.

The predicted number counts are shown in Fig.~4 together with
the actual GRB {\it SWIFT} counts. A meaningful comparison can
only consider `bright' {\it SWIFT} GRBs, namely above a photon
flux of $1$ ph s$^{-1}$ cm$^{-2}$, as below this level some
degree of incompleteness is expected (Band 2006).

\begin{figure}
\includegraphics[width=\columnwidth]{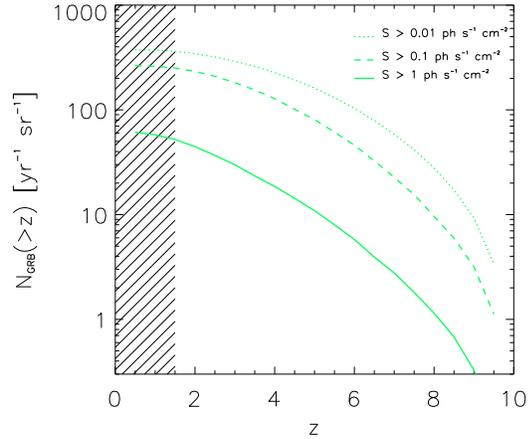}
\caption{The redshift distribution of GRBs at the limiting flux
[15-150 keV] of $0.01$ ({\it dotted} line), $0.1$ ({\it dashed} line),
and $1$ ph s$^{-1}$ cm$^{-2}$ ({\it solid} line), computed with a
threshold on metallicity at $Z_{\odot}/10$.}
\end{figure}

\begin{figure}
\includegraphics[width=\columnwidth]{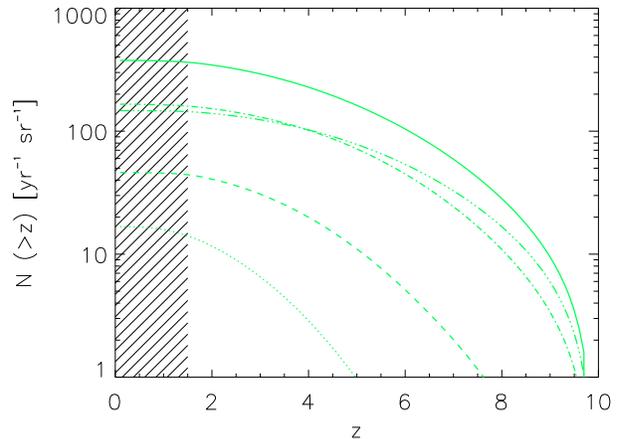}
\caption{The redshift distribution of GRBs computed with a threshold
on metallicity at $Z_{\odot}/10$ ({\it solid} line).  The other curves
illustrate the contributions from host galaxies with different halo
masses: $10^9-10^{10}\, M_{\odot}$ ({\it triple-dot-dashed} line),
$10^{10}-10^{11}\, M_{\odot}$ ({\it dot-dashed} line),
$10^{11}-10^{12}\, M_{\odot}$ ({\it dashed} line), $10^{12}-10^{13}\,
M_{\odot}$ ({\it dotted} line).}
\end{figure}

As it is apparent from Fig.~4 a reasonable agreement with the
{\it SWIFT} counts can be obtained after the above assumptions.
It should be pointed out that we did not perform any fitting
optimization, but simply compared the counts predicted by the
model with the data, under the simplified and commonly adopted
assumptions on the luminosity and spectral energy GRB
distributions outlined above.

A very interesting aspect concerns the fact that the counts can
be reasonably reproduced without requiring any GRB prompt
luminosity evolution (e.g. Daigne et al. 2006). This is a
natural consequence of the fact that our redshift progenitor
distribution intrinsically peaks at redshift higher than the
cosmic star formation.  Indeed, if we were to reproduce the
number counts with no metallicity threshold, the number of GRBs
per unit mass of formed stars ($k$) would have to be a factor
about $10$ lower, which would imply that the bulk of the
progenitors would be located at lower redshift.

\begin{figure}
\begin{center}
\includegraphics[width=6truecm, height=16 truecm]{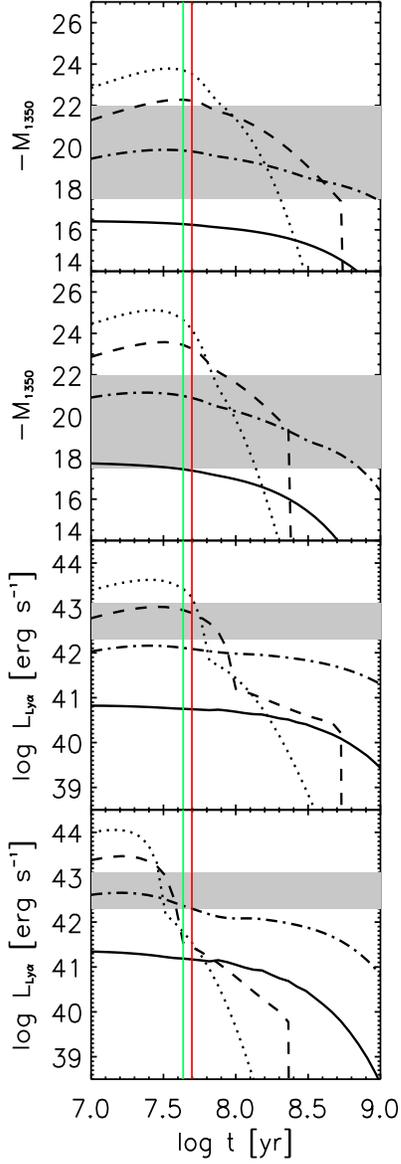}
\caption{Extincted magnitude at $1350$ \AA~ and Ly$\alpha$ luminosity
of the host galaxies of the GRB progenitors as a function of galactic
age. Lines as in Fig.~1. The first and third panels refer to redshift
$3$, the second and forth ones to redshift $6$. The \textit{shaded}
areas illustrate the ranges sampled in the observed luminosity
functions.}
\end{center}
\end{figure}

The predicted redshift distribution of GRB events for different
flux limits is presented in Fig.~5. This shows that already at
fluxes $S\ga 1$ ph s$^{-1}$ cm$^{-2}$ we expect that $10$ per
cent of GRBs occur at $z \ga 6$. Of course such a fraction
increases with decreasing limiting flux, reaching around $30$
per cent for $S\ga 10^{-2}$ ph s$^{-1}$ cm$^{-2}$; since this
flux limit is below the present detectable level, the upper
curve of Fig.~5 represents the redshift distribution of all
currently observable GRBs.

More specifically, the model predicts that in two years 13 GRBs
at $z\ga 6$ and flux limit $1$ ph s$^{-1}$ cm$^{-2}$ should
have been detected by {\it SWIFT}. So far only 1 GRB has robust
redshift estimate at $z>6$. On the other hand, while it is
difficult to quantify the efficiency of the redshift
determination, especially for very high redshift events, this
could well be less than $10$ per cent.

\subsection{GRB host galaxies}

The exploited galaxy formation model also enables us to
directly predict the properties of the GRB host galaxies (SFR,
magnitude, stellar mass, average metallicity, extinction) as
function of the halo mass and age as well as the luminosity
function (LF) and the corresponding SFR distribution of the
overall population.

As discussed in \S~2.1, during the evolution of individual
galaxies, the metallicity of the star-forming gas attains the
threshold $Z_{\rm crit}=0.1-0.2 \, Z_{\odot}$ quite rapidly,
within $t_{\rm crit}\approx 5\times 10^7$, and this timescale is
basically independent of the host halo mass.

Interestingly, since this is much shorter than the halo
virialization timescale, $t_{\rm crit} \ll t_{\rm H}$, the GRB
rates for fixed halo mass directly trace the cosmic rate of
halo virialization. As a consequence, the GRB hosts mostly
reside within galaxy halos in the mass range $10^9\la M_{\rm
H}\la 10^{11}\, M_{\odot}$, and only at $z\la 4$ the fraction
of hosts in massive halos $M_{\rm H}\ga 10^{11}\, M_{\odot}$
exceeds $10$ per cent. The relative number of GRBs for
different halo masses as function of redshift is shown in
Fig.~6.

The predicted evolution of the SFR and stellar mass as
functions of galactic age and halo mass are reported in Fig.~1.
For halo masses in the range $10^{10}-10^{13}\, M_{\odot}$ and
virialized at redshift $z\approx 6$ the expected SFR spans
$0.3-10^3\, M_{\odot}$ yr$^{-1}$, and the corresponding stellar
masses cover the interval $10^7\la M_{\star} \la 2\times
10^{10}\, M_{\odot}$, with both quantities scaling as
$(1+z)^{3/2}$ at a given halo mass. The shortness of $t_{\rm
crit}$ yields high specific SFRs, $\dot{M_{\star}}/M_{\star}\ga
2 \times 10^{-8}$, almost independently of the virialization
redshift.

Now we turn to consider more directly observable properties of
GRB hosts, such as the average ultraviolet extinction
$A_{1350}$ and the corresponding extincted magnitude $M_{1350}$
at $1350$ \AA, again as function of galactic age. We recall
that $M_{1350}\approx
-18.6-2.5\,\log(\dot{M}_{\star}/M_{\odot}~\mathrm{yr}^{-1}) +
A_{1350}$ (see Appendix A). Fig.~1 reports $A_{1350}$, which
scales with redshift as $(1+z)^{3/5}$, for various halo masses
virialized at redshift $z=6$. A key prediction of the model is
apparent, namely that most GRB hosts are poorly extincted
systems. Since the less massive hosts ($M_{\rm H}\la 10^{11}\,
M_{\odot}$) largely outnumber the most massive ones, the
typical $A_{1350}$ ranges between $0.01 - 0.3$ mag,
corresponding to $A_{\rm V}\la 0.1$. Larger dust extinction is
predicted only for more massive hosts $M_{\rm H}>10^{11}\,
M_{\odot}$, which exceed a few per cent of the total GRB hosts
only at $z\la 6$ (cf. Fig.~1).

The AB absolute magnitude at 1350 {\AA}, $M_{1350}$, is
reported in Fig.~7 for various halo masses and redshifts $z=3$
and $z=6$.  The shaded area indicates the absolute magnitude
range where the UV high-$z$ LF is currently well sampled. It is
clear that the UV luminosity is practically not affected by
dust, since the predicted extinction is low for galaxy ages
less than $t_{\rm crit}$.

Fig.~7 also shows the age dependence of the expected Ly$\alpha$
luminosity at fixed halo mass. At variance with respect to the
behavior of the UV magnitude, the Ly$\alpha$ luminosity of
hosts in larger halos is already declining for ages less than
$t_{\rm crit}$ at $z\ga 6$, since Ly$\alpha$ emission is
sensitive not only to dust but also to neutral hydrogen
absorption.

\begin{figure*}
\includegraphics[height=6truecm]{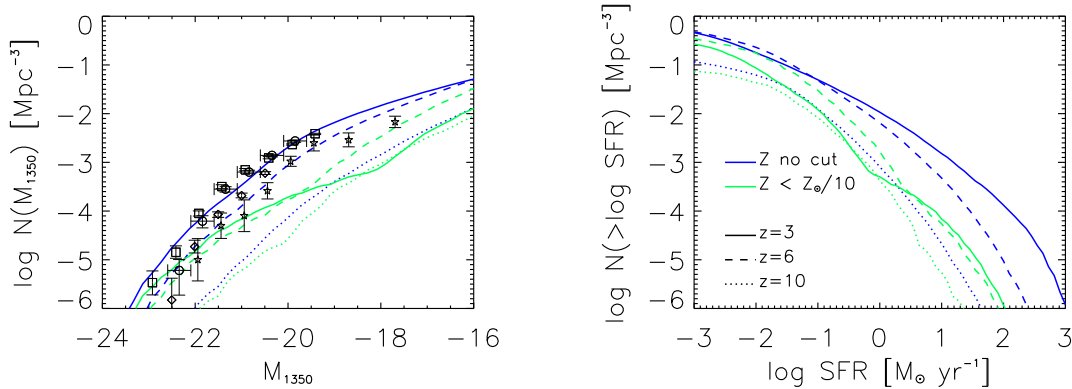}
\caption{The comoving number density of galaxies as a function
of UV magnitude at $1350$ \AA\, $M_{1350}$ (\textit{left}) and
of SFR (\textit{right}). In both panels \textit{solid},
\textit{dashed} and \textit{dotted} lines refer to redshifts
$z=3$, $6$, and $10$, respectively. The sets of \textit{blue}
and \textit{green} curves refer to no metallicity threshold and
$Z_{\rm crit} = Z_{\odot}/10$. The data points reported in the
left hand panel are from Steidel et al. (2001;
\textit{circles}) and Yoshida et al. (2006; \textit{squares})
at $z\sim 3-4$, Yoshida et al.  (2006; \textit{diamonds}) at
$z\sim 4-5$, and Bouwens et al.  (2006; \textit{stars}) at
$z\sim 5-7$.}
\end{figure*}

Anyhow we expect a strict correlation between GRB hosts and
Lyman Break Galaxies (LBG) and Ly$\alpha$ emitters (LAE).  As
the galaxy model here adopted reproduces the high-redshift UV
LF of LBGs and LAEs (see Mao et al. 2007), it is meaningful to
estimate the UV LF expected for GRB hosts, namely by imposing
the condition $Z\la Z_{\rm crit}\approx 0.1\, Z_{\odot}$. The
result is shown in Fig.~8. In general the conditional UV LF
depends on the ratio between $t_{\rm crit}$ and the time during
which the host is brighter than a fixed absolute magnitude. As
expected, the metallicity cut depresses the UV luminosity
distribution, but this depression is minimal at the faint and
bright ends, and maximum at intermediate values, as expected
from the effects of the cut on $M_{1350}$ in individual
systems.

In particular, the time spent by galaxies in large halos as
LBGs - before turning into strongly submm emitting galaxies
because of dust attenuation - is short and comparable with
$t_{\rm crit}$.  At the faint end the luminosity of small hosts
can fall below a fixed magnitude in a relatively short time, as
shown in Fig.~8 for hosts in halos with $M_{\rm H}\approx
10^{10}\, M_{\odot}$. It is also apparent that intermediate
mass halos and the corresponding intermediate UV luminosity
objects are significantly more affected by the condition on
metallicity. Obviously at high redshift, e.g. $z\approx 10$,
the cut reduces the visibility timescale only by a small
fraction.

We can conclude that GRB hosts should well reproduce the LBG LF
at its faint and bright ends. Similar conclusions hold for LAE
LF.

The shortness of $t_{\rm crit}\approx 50$ Myr implies that the
hosts must exhibit quite large [$\alpha$/Fe] enhancement. In
fact, the cumulative fraction of SNIa explosions after an
instantaneous burst of star formation ($1$ at $12$ Gyr) is
negligibly small for $50$ Myr (corresponding to the lifetime of
a $7\, M_\odot$ star), for a wide class of progenitor models,
i.e. Single Degenerates (SD), and Double Degenerates (DD)
exploders (Greggio 2005). At $0.1$ Gyr it reaches $10$ per cent
in the most favorable case of close DD scheme.  Furthermore,
the typical timescales required to significantly decrease the
[$\alpha$/Fe] ratio from its initial value produced by a single
generation of massive stars (approximately $0.4-0.6$ depending
on metallicity, IMF and stellar yields, Gibson et al. 2003), is
$0.3$, $1$ and $3$ Gyr, for close DD, SD and WIDE DD
progenitors, respectively. Thus while in the most massive
ellipticals the duration of the burst of star formation (around
$0.3$ Gyr) may be enough for a mild pollution by Fe-peak
elements (depending on the assumed scenario for SNIa
progenitors), the lower timescale required to reach the
critical metallicity implies that GRB hosts should display the
original pattern of heavy elements produced by massive star
chemical evolution. Notice that the subsequent history of star
formation decreases the initial $\alpha$-enhancement more in
smaller than in larger objects (see Fig.~1). We conclude that
at high redshift, GRB hosts (and the similarly young LAEs)
exhibit the highest values of $\alpha$-enhancement, even higher
than those of high-$z$ quasar hosts and of their descendents,
the massive elliptical galaxies.

Finally, we can also estimate the distribution of intrinsic
SFRs. It is worth stressing that generally the hosts with SFRs
exceeding a few $\times 10^2\, M_{\odot}$ yr$^{-1}$ are
significantly less than the overall galaxy population. This
suppression reflects not only the steep halo mass function, but
also the short duty cycle of GRB hosts following the rapid
metal enrichment in the initial stages of star formation. And
again the effect gets more pronounced at decreasing $z$ due to
the increased fraction of observable hosts which are below
threshold. On the other hand, the model allows for the
existence of GRB hosts with SFRs exceeding a few $\times 10^2\,
M_{\odot}$ yr $^{-1}$ within halos endowed with $M_{\rm H}\ga
10^{12}\, M_{\odot}$: as shown in Fig.~1 and Fig.~8, the
fraction of these hosts should amount to a few per cent of the
total. Nevertheless they should not have formed a large amount
of stars $M_{\star}\la 10^{10}\, M_{\odot}$, exhibit specific
SFR $\dot{M_{\star}}/M_{\star}\ga 10^{-8}$ yr$^{-1}$ and should
be rather extincted by dust (see Fig.~1).

\section{Discussion}

The adopted galaxy formation scenario coupled with the metal
poor collapsar model suggested by stellar evolution have been
exploited to infer the above results, which include GRB counts
and redshift distribution and the complete description of the
relevant properties of their hosts, such as SFR, mass in stars,
chemical evolution in the cold star forming gas and in the
stellar component within individual galaxy halos, with
specified mass and formation redshift.  In this \S~4 these
results are discussed in the light of presently available
observations and they are compared to previous studies.

\subsection{GRB progenitor rates and GRB counts}

The imposed metallicity threshold affects the GRB progenitor
redshift distribution. Instead of peaking at around $z\approx
3$ as in the case without metallicity constrain, the threshold
$Z_{\rm crit}=Z_{\odot}/10$ yields a broad peak at $z\approx 6$
(Fig.~2). A significant fraction of progenitors are expected at
high redshift: approximately $60$ per cent at $z\ga 4$ and
approximately $30$ per cent at $z\ga 6$ (see Fig.~3). This
behavior is shared by all models which associate GRBs with
metal poor progenitors (e.g. Natarajan et al. 2005; Langer \&
Norman 2006; Salvaterra \& Chincarini 2007).

In particular, our findings are similar to those inferred by
Langer \& Norman (2006), who explored the effect of a
metallicity threshold in the environment of GRBs, by adopting a
mass-stellar metallicity relation for galaxies and an average
cosmic metallicity $[Z]$ dependence on redshift ${\rm
d}[Z]/{\rm d}z\approx -0.15$ dex. This law, derived to
reproduce the metallicity of stars in galaxies, has been
extrapolated to the ISM. This extrapolation implies that at
$z\ga 7$ all cosmic star formation occurs in environments with
$[Z]\la 0.1\, Z_{\odot}$, independently of the processes
occurring in galaxy halos. As a result the corresponding rate
of low metallicity SNae peaks at $z\approx 6$. In our model the
combination of cooling and feedback processes implies that at
$z\approx 7$ the SFR in cold gas with $Z\la 0.1\, Z_{\odot}$
amounts to about $40$ per cent of the cosmic one, due to the
short timescale for gas enrichment. Therefore, despite of the
quite different approach, the redshift distributions of GRB
progenitors found by Langer \& Norman (2006) and Yoon et al.
(2006) are very similar to ours, since their assumed SFR for
metal poor stars peaks at $z\approx 6$, similarly to our SFR in
cold gas environment (see Fig.~2).

The combination of the galaxy formation scenario with the low
metallicity collapsar hypothesis for GRB events leads to a good
agreement of the predicted and observed (bright, $S\ga 1$ ph
s$^{-1}$ cm$^{-2}$) {\it SWIFT} counts (see Fig.~4). This
naturally follows - without any tuning - from the derived
number of GRB progenitors, under quite simple and plausible
assumptions on the GRB jet opening angle, prompt gamma-ray
luminosity distribution and spectral shape, without requiring
any luminosity evolution (see \S~2.3 and \S~3.2).

The corresponding redshift distribution (see Fig.~5) implies
that at this bright limit the adopted model predicts about $6$
GRBs per year, while only one GRB at $z\ga 6$ has been
identified in two years of {\it SWIFT} operation.  As a matter
of fact, only for about half of the {\it SWIFT} bursts an
optical afterglow has been observed, and for only $30$ per cent
it has been possible to infer a redshift estimate.  It is
therefore quite reasonable to guess that this fraction could be
about $10$ per cent for bursts at $z\ga 6$ (see Fiore et al.
2007).

Recently Salvaterra \& Chincarini (2007) obtained a good formal
fit to the {\it SWIFT} counts down to $S\approx 0.4$ ph
s$^{-1}$ cm$^{-2}$, adopting as free parameters the count
normalization and the GRB LF (two further parameters).  They
considered the case of low metallicity environment by adopting
a kinematical model and predict the occurrence of only 1 GRB
yr$^{-1}$ for the {\it SWIFT} bright flux limit ($1$ ph
s$^{-1}$ cm$^{-2}$). Their result imply a redshift
determination efficiency for GRBs greater than $50$ per cent.

It is worth noticing that reaching completeness down to $0.1$
ph$^{-1}$ s$^{-1}$ cm$^{-2}$ would significantly increase the
number of detected GRBs and in turn allow to explore the
Universe during the recombination epoch, $z \ga 8$, with good
statistical significance. A further decrease in the flux limit
to $0.01$ ph$^{-1}$ s$^{-1}$ cm$^{-2}$, would instead only
increase the GRB sample by a factor of $2$; at this flux limit
practically all GRBs would be detected (see Figs.~5 and 6).

In order to account for the trend of GRBs to be at substantial
redshift Firmani et al. (2005) proposed that the GRB LF is
evolving. Daigne et al. (2006) tested the hypothesis of an
increasing efficiency of GRB production by massive stars with
increasing redshift. Detailed redshift distributions of GRBs
will discriminate among these different possibilities.

We can conclude that the hypothesis that metal-poor, rapidly
rotating massive stars are the GRB progenitors (Woosley \&
Heger 2006; Yoon et al. 2006) is consistent with the observed
{\it SWIFT} counts. Clearly determinations of GRB redshifts
will be extremely informative on the progenitor and galaxy
formation models.

We stress that the adopted galaxy formation scenario exploits
quite a standard IMF, independent of the gas metallicity. As a
matter of fact, we showed that the cold star forming gas is
rapidly ($t\ll 10^7$ yr) enriched to the possible threshold
around $3-5 \times 10^{-4}\, Z_{\odot}$, below which the IMF
might be strongly biased toward high mass stars (see Bromm et
al. 2001; Schneider et al. 2006). Therefore the possible
contribution of pop III stars with IMF strongly biased toward
high masses is not considered here. However, Bromm \& Loeb
(2006) showed that at $z\approx 10$ the contribution of pop III
to cosmic SFR could be of order of 1/10 of the overall SFR,
becoming dominant at $z\ga 15$; the GRB rate from pop III
massive stars would grow correspondingly.

\subsection{Properties of Long GRB Host Galaxies}

Though the idea that GRBs are preferentially located in metal
poor environments is attractive, nevertheless observational
estimates of the metal content of host galaxies are still
controversial. While most of the results suggest that at
high-$z$ GRB hosts exhibit metallicity $Z\la 0.1-0.3\,
Z_{\odot}$ (Prochaska et al. 2007, Price et al. 2007), there
are claims of higher metal content (see Savaglio et al. 2003).

Once a metal poor ($Z\sim 0.1\, Z_{\odot}$) collapsar model is
assumed, our galaxy formation scenario predicts that GRB hosts
are very young, with age less than $5 \times 10^{7}$ yr,
independently of the halo mass. This young age directly mirrors
the predicted short timescale, $t_{\rm crit}\approx 5\times
10^{7}$ yr, of chemical enrichment of the cold gas,
independently of mass. Such independence makes the GRB rate at
high-$z$ a good tracer of the virialization rate of relatively
small galaxy halos, $M_{\rm H}\la \times 10^{11}\, M_{\odot}$
(see Fig.~6).

A definite prediction issuing from their youth is that GRB
hosts should exhibit high [$\alpha$/Fe]-enhancement, as their
metal content directly reflects the chemical yields of core
collapse SNae.  Indeed ,in the most favorable scenario of close
DD progenitors, type Ia SNae may halve the [$\alpha$/Fe] ratio
produced by the generations of massive stars in a few 10$^8$
yr, somewhat longer than $t_{\rm crit}$. Recent observations by
Prochaska et al. (2007) suggest that $\alpha$/Fe ratios are
more than $3$ times the solar value. Though differential
depletion could be responsible for the result, the young age of
the hosts is a much more palatable explanation.

The shortness of $t_{\rm crit}$ implies that star formation has
not much proceeded, i.e. the model predicts high specific star
formation $\dot{M_{\star}}/M_{\star}\ga 2 \times 10^{-8}$
yr$^{-1}$. Notice that even the specific star formation is
almost independent of halo mass. Specific SFRs at this high
level have been claimed by several authors (Fruchter et al.
1999; Le Floc'h et al. 2003; Christensen et al. 2004; Savaglio
et al. 2006; Castro Cer\'on et al. 2006; Micha{\l}owski et al.
2008). We caution that metal poor gas may be left over
particularly in external regions of relatively old galaxies,
hosting a burst of star formation even if most stars already
formed and the gas was already metal enriched. However, we
expect that relatively evolved hosts endowed with significant
stellar mass are exceptions: GRB~020127 could be one of such
cases (Berger et al. 2007). The vast majority of GRBs are
hosted by small galaxies soon after their first stars shine.

Since the hosts are very young galaxies, our model predicts
that they are gas rich objects with large column density
$N_{\rm H}$, once more independently of their galaxy halo mass.
Observations confirms the tendency for GRB hosts to exhibit
large $N_{\rm H}$ (Prochaska et al. 2007; Schady et al. 2007).
At odd with these results, Tumlinson et al. (2007) set
stringent upper limits on the molecular hydrogen ($H_2$)
abundance, concluding that the fraction of molecular hydrogen
$H_2/HI$ is extremely low in the examined hosts. These authors
also point out that the deficiency may be related to low dust
abundance, $H_2$ formation being catalyzed on the surface of
dust grains. An additional possibility is the destruction of
$H_2$ by UV radiation. The model here proposed predicts that
GRB hosts are dust poor and are pervaded by a intense UV
radiation field; therefore they are expected to be quite poor
in molecular hydrogen.

Concerning the UV emission, most of the hosts ($M_{\rm H}\la
10^{10}\, M_{\odot}$) should have UV magnitude at $1350${\AA}
in the range $-20\la M_{1350}\la -16$. The most luminous hosts
are bright enough to be included in the already available LBG
LF.  Interestingly, Jakobsson et al. (2005) have shown that GRB
hosts are tracing the faint end of the LBG LF at $z\approx 3$.

The predicted host Ly$\alpha $ luminosity should fall in the
interval $2\times 10^{40}\la L_{\rm Ly\alpha}\la 2\times
10^{42}$ erg s$^{-1}$, only marginally overlapping with the
range explored by currently available Ly$\alpha$ LF (Fig.~7).
Tanvir \& Levan (2008) have shown that the UV rest frame
luminosity distribution of Ly$\alpha$ selected galaxies and GRB
hosts at $z\sim 3$ are quite similar.  We notice that Mao et
al. (2007) pointed out that LAE are expected to be younger,
with lower stellar masses, more compact and associated with
less massive halos than LBGs.

In summary, observations supports the model prediction that the
GRB hosts trace the faint end of the LF of LBGs and LAEs.

A further issue concerns the amount of dust in GRB hosts. While
most hosts have not been detected at mid-IR or submm
wavelengths (e.g. Tanvir et al. 2004; Le Floc'h et al 2006;
Priddey et al. 2006) in a few cases submm and radio emission
have been actually detected (Micha{\l}owski et al. 2008).  How
can the existence of these hosts be interpreted within the
proposed scenario?

The model predicts that only GRBs hosted in large galactic
halos $M_{\rm H}\ga 10^{12}\, M_{\odot}$ can have significant
dust absorption. These also exhibit large SFR
($\dot{M_{\star}}\ga 100\, M_{\odot}$ yr$^{-1}$) and relatively
small stellar mass ($M_{\star}\la 10^{10}\, M_{\odot}$). These
properties correspond to those inferred by Micha{\l}owski et
al. (2008) for the submm and radio detected objects. The model
also predicts that these systems should represent only $1/20$
of all the GRB hosts at $z\approx 1-2$ and practically
disappear at $z \ga 5$.

We stress that the conclusions regarding the host properties
refer to the average GRB population, and do not exclude the
possibility that individual GRBs might reside in low
metallicity local regions within their host (as might be the
case for GRB~060206, Fynbo et al. 2006).

The above findings have been derived in the framework of a
galaxy formation scenario, where a key role is played by the
energy feedbacks provided by SNae and quasars. However, the
fast chemical enrichment leading to gas metallicity above
$Z_{\rm crit}\approx 0.1\, Z_{\odot}$ takes place at early
galactic times, before quasar feedback becomes effective. On
the other hand, the stellar feedback is very relevant since it
regulates the star formation activity even at early galactic
times.

\section{Summary}

We have explored the cosmological consequences of the
assumption that metal poor and rapidly rotating single stars
are the progenitors of most long GRBs. Our main conclusions
are:

$\bullet$ The overall long GRB rate amounts to approximately
$300$ yr$^{-1}$ sr$^{-1}$. Bright {\it SWIFT} counts are
reproduced by assuming a non-evolving prompt (gamma-ray)
luminosity distribution.

$\bullet$ Above a flux limit of $1$ ph s$^{-1}$ cm$^{-2}$ about
$30$ per cent of GRBs are predicted to be at $z\ga 6$ and $10$
per cent at $z\ga 8$, amounting to approximately $13$ for two
years of {\it SWIFT} operation.  Only one have been located
above $z\ga 6$ in two years: this would require a redshift
determination efficiency around $10$ per cent, to be compared
with the $30$ per cent efficiency for events at lower $z$.

$\bullet$ The host galaxies are very young, with age less than
$5\times 10^{7}$ yr, gas rich (large column densities) but
poorly extincted systems ($A_{\rm V}\la 0.1$), because of their
chemical immaturity. Only the few per cent of hosts associated
with large halos ($M_{\rm H}\ga 10^{12}\, M_{\odot}$) have
large extinction ($A_{\rm V}\ga 0.3$), high SFR
($\dot{M_{\star}}\ga 100\, M_{\odot}$ yr$^{-1}$) and can be
detected at relatively bright submm flux levels. This result
has implications for the origin of `dark' GRBs [about $1/3$ of
{\it SWIFT} GRBs can be considered `dark', e.g. Schady et al.
(2007)], lacking a detection of the optical afterglow. Dark
GRBs should largely comprise a population of high $z$ events,
rather than highly extincted systems.

$\bullet$ The young age of hosts implies that (i) the specific
SFR is high ($\dot{M_{\star}}/M_{\star}\ga 2 \times 10^{-8}$)
and (ii) the ratios of abundance of different chemical elements
are just those of the respective chemical yields of type II
SNae, i.e. large $\alpha$-enhancements should be the rule.

$\bullet$ Most of the hosts ($10^{9}\la M_{\rm H}\la
10^{11}\,M_{\odot}$) have UV magnitude in the range $-20\la
M_{1350}\la -16$ and Ly$\alpha$ luminosity in the range
$2\times 10^{40}\la L_{\rm Ly\alpha}\la 2\times 10^{42}$ erg
s$^{-1}$. They trace the formation of small galaxies in small
halos, and as a consequence the faint end of the LBG and LAE
LF. These hosts would reionize the Universe at $z\approx 7$
(see Mao et al. 2007).

\section*{Acknowledgments}
We thank Giancarlo Ghirlanda for helpful discussions,
Gianfranco De Zotti for useful suggestions and critical reading
of the manuscript, and the anonymous referee for an
illuminating report. This research is partially supported by
MIUR, INAF and the NSF under Grant No. PHY99-07949. AC thanks
the KITP (Santa Barbara) for kind hospitality.

\begin{appendix}

\section{A simple recipe for star formation and dust obscuration
in protogalaxies}

Granato et al. (2004) have proposed a model for early galaxy
formation, in which the most relevant processes (gas cooling
and inflow, star formation and gas accretion onto BH, stellar
and quasar feedback, gas outflow) are included in a set of
equations, that can be solved with straightforward numerical
computations. In this Appendix we present a simplified version
of the model describing the SFR, mass in stars and chemical
evolution, which are relevant for this work. The analytical
formulae presented below are very good approximations of the
results found by solving the full set of equations of the model
(for details see Lapi et al. 2006 and Mao et al. 2007). We
stress that, because of their fundamental character, the
equations listed below describes the main aspects of star
formation and chemical evolution in protogalaxies at high
redshift.

When a DM halo of mass $M_{\rm H}$ reaches the virial
equilibrium, it contains a mass
$M_{\mathrm{inf}}(0)=f_{\mathrm{cosm}}\,M_{\rm H}$ of hot gas
at the virial temperature, $f_{\mathrm{cosm}}\approx 0.18$
being the mean cosmic baryon to DM mass-density ratio. The gas
in virial equilibrium flows toward the central region at a rate
$\dot{M}_{\mathrm{cond}}=M_{\mathrm{inf}}/t_{\mathrm{cond}}$
where the {\it condensation} timescale
$t_{\mathrm{cond}}=\max[t_{\mathrm{cool}}(R_{\rm
H}),t_{\mathrm{dyn}}(R_{\rm H})]$, is the maximum between the
dynamical time and the cooling time at the halo virial radius
$R_{\rm H}$. When computing the cooling time, a clumping factor
${C}$ in the baryonic component is also included: ${C}\ga 7$
implies $t_{\mathrm{cool}}(R_{\rm H})\la
t_{\mathrm{dyn}}(R_{\rm H})$ on relevant galaxy scales at
high-$z$. By defining such a condensation time, we implicitly
neglect the effect of angular momentum. However, angular
momentum decays on a dynamical friction timescale
$t_{\mathrm{DF}}\approx 0.2\, (\xi/\ln{\xi})\,
t_{\mathrm{dyn}}$, where $\xi=M_{\rm H}/M_{\rm c}$ and $M_{\rm
c}$ is the typical mass cloud involved in major mergers (e.g.
Mo \& Mao 2004); major mergers, which are very frequent at high
redshift and in the central regions of halos, imply $\xi \sim$
a few.

The model also assumes that quasar activity removes the hot gas
from the halo through winds at a rate
$\dot{M}_{\mathrm{inf}}^{QSO}$; the equation for the diffuse
hot gas is then
\begin{equation}
\dot{M}_{\mathrm{inf}}=
-\dot{M}_{\mathrm{cond}}-\dot{M}_{\mathrm{inf}}^{QSO}~.
\end{equation}

The cold gas is piled up following the cooling of hot gas, is
consumed by star formation ($\dot{M}_{\star}$), and is removed
by the energy feedback from SNae
($\dot{M}_{\mathrm{cold}}^{SN}$) and quasar activity
($\dot{M}_{\mathrm{cold}}^{QSO}$):
\begin{equation}
\dot{M}_{\mathrm{cold}} = \dot{M}_{\mathrm{cond}}-
(1-\mathcal{R})\dot{M}_{\star} -
\dot{M}_{\mathrm{cold}}^{SN}-\dot{M}_{\mathrm{cold}}^{QSO}~,\\
\end{equation}
where $\mathcal{R}$ is the fraction of gas restituted to the
cold component by the evolved stars. Under the assumption of
instantaneous recycling, $\mathcal{R}\approx 0.54$ for the
adopted IMF ($\mathcal{R}\approx 0.3$ for a Salpeter IMF); this
value of $\mathcal{R}$ is an upper limit, since only a fraction
of evolved stars have a significant mass loss in the
evolutionary phases considered here. However, the relevant
results are only very weakly sensitive to the chosen value, in
the physically allowed range. The mass of cold baryons that is
going to be accreted onto the central supermassive BH is small
enough to be neglected in the above equation (see Granato et
al. 2004).

Stars are formed at a rate
\begin{equation}
\dot{M_{\star}}=\int
\frac{\mathrm{d}M_{\mathrm{cold}}}{\max[t_{\mathrm{cool}},t_{\mathrm{dyn}}]}
\approx \frac {M_{\mathrm{cold}}}{t_{\star}}~,
\end{equation}
where now $t_{\mathrm{cool}}$ and $t_{\mathrm{dyn}}$ refer to a mass
shell $\mathrm{d}M_{\mathrm{cold}}$, and $t_{\star}$ is the star
formation timescale averaged over the mass distribution.

The rate of cold gas removal due to SNae is parameterized as
\begin{equation}
\dot{M}_{\mathrm{cold}}^{SN} = \beta_{\rm SN}\, \dot{M}_{\star}~,
\end{equation}
where the averaged efficiency
\begin{equation}\nonumber
\beta_{\rm SN}=\frac {n_{\rm SN}\,
\epsilon_{\rm SN}\,E_{\rm SN}}{E_{\mathrm{bind}}}\approx 0.6
\left(\frac{n_{\rm SN}}{1.4\times 10^{-2}/ M_{\odot}}\right)\times
\end{equation}
\begin{equation}
\left(\frac{\epsilon_{\rm SN}}{0.05}\right)
\left(\frac{E_{\rm SN}}{10^{51}\mathrm{erg}}\right)
\left(\frac{M_{\rm H}}{10^{12} M_{\odot}}\right)^{-2/3} \left(
\frac{1+z}{7}\right)^{-1}
\end{equation}
depends on the number of SNae per unit solar mass of condensed
stars $n_{\rm SN}$, the energy per SN available to remove the
cold gas $\epsilon_{\rm SN}\,E_{\rm SN}$, and the specific
binding energy of the gas within the DM halo,
$E_{\mathrm{bind}}$. Following Zhao et al. (2003) and Mo \& Mao
(2004), the latter quantity has been estimated, for $z\ga 1$,
as $E_{\mathrm{bind}}=V_{\mathrm{H}}^{2}\,
f(c)\,(1+f_{\mathrm{cosm}})/2\approx 5.6\times 10^{14}\,
(M_{\rm H}/10^{12}\, M_{\odot})^{2/3}\, [(1+z)/7]\,
{\mathrm{cm}^2~\mathrm{s}^{-2}}$. Here $V_{\rm H}$ is the halo
circular velocity at the virial radius and $f(c)\approx
2/3+(c/21.5)^{0.7}\sim 1$ is a weak function of the halo
concentration $c\sim$ a few. Lapi et al. (2006) have shown that
high redshift LFs of quasars and galaxies constrain
$\epsilon_{\rm SN}\approx 0.05$; the same value is required in
order to reproduce the fundamental correlations between local
ellipticals and dormant BHs.

By analyzing the results of the numerical solution of the full
set of equations by Granato et al. (2004), it is apparent that
the term of quasar feedback is important only during the final
stage of BH growth, around $2-3$ e$-$folding times
(approximately $10^8$ yr) before the peak of quasar luminosity,
when the energy discharged by the quasar is so powerful to
unbind most of the residual gas, quenching both star formation
and further accretion onto the supermassive BH.  The time
integral over the quasar bolometric power exceeds the gas
binding energy after
\begin{equation}
\Delta t_{\mathrm{burst}}\approx 2.5\times 10^8 \,
\left(\frac{1+z}{7}\right)^{-1.5}\, \mathcal{F}\left(\frac
{M_{\rm H}}{10^{12}\, M_{\odot}}\right)~~\mathrm{yr}~,
\end{equation}
where $\mathcal{F}(x)=1$ for $x\ga 1$ and
$\mathcal{F}(x)=x^{-1}$ for $x\la 1$. Therefore a good
approximation for the star formation history is obtained by
neglecting the quasar feedback effect in Eqs.~(A1) and (A2) and
by abruptly stopping star formation and accretion onto the
central BH after$\Delta t_{\mathrm{burst}}$ since halo
virialization.

Then Eqs.~(A1) and (A2) can be easily solved, with the outcome
that the infalling mass declines exponentially as
$M_{\mathrm{inf}}(t) = M_{\mathrm{inf}}(0)\,
\mathrm{e}^{-t/t_{\mathrm{cond}}}$, while the SFR evolves
according to
\begin{equation}
\dot{M}_{\mathrm{\star}}(t)=\frac
{M_{\mathrm{inf}}(0)}{t_{\mathrm{cond}}(\gamma-1/s)} \left[
\mathrm{e}^{-t/t_{\mathrm{cond}}}- \mathrm{e}^{-s\,\gamma\,
t/t_{\mathrm{cond}}}\right]~,
\end{equation}
with $\gamma\equiv 1-\mathcal{R}+\beta_{\rm SN}$.  The quantity
$s\equiv t_{\mathrm{cond}}/t_{\mathrm{\star}}$ is the ratio between
the timescale for the large-scale infall estimated at the virial
radius and the star formation timescale in the central region; it
corresponds to $s\sim5$, both for an isothermal or NFW (Navarro et
al. 1997) density profile.

The following expression well approximates the condensation
timescale (see Mao et al. 2007):
\begin{equation}
t_{\mathrm{cond}}\approx 4\times 10^8\,
\left(\frac{1+z}{7}\right)^{-1.5}\, \left(\frac {M_{\rm H}}{10^{12}\,
M_{\odot}}\right)^{0.2}~~\mathrm{yr}~.
\end{equation}
The scaling with redshift reflects the dependence of the
dynamical time; the weak dependence on $M_{\rm H}$ reproduces
the impact of the energy feedback from the quasar on the
infalling gas, which is stronger for more massive halos hosting
more massive BH.

In order to compute the metal content of the cold gas, one has
to take into account the infall of primordial abundance gas,
the enrichment due to earlier generations of stars, and
outflows driven by winds generated by SNae and quasars. We
assume, as common, that in the cold gas there is an
instantaneous mixing of metals released by stars. The
corresponding simplified equation reads
\begin{equation}
\dot{Z}_{\mathrm{cold}}(t)=\frac {Z_{\mathrm{inf}}-Z_{\mathrm{cold}}
}{t_{\mathrm{cond}}} \frac
{M_{\mathrm{inf}}(t)}{M_{\mathrm{cold}}(t)}+ \frac
{\mathcal{R}_{\mathrm{Z}}(t)}{t_{\star}},
\end{equation}
where
\begin{equation}
{\mathcal{R}_{\mathrm{Z}}(t)}\approx
\int_{m_{\star,t}}^{m_\star^{\rm sup}}{\rm d}m_{\star}\,
m_{\star}\, q_Z(m_{\star})\, \phi(m_{\star})\,
{\dot{M}_{\star}(t-\tau_{m_\star})\over \dot{M}_{\star}(t)}~.
\end{equation}
The IMF is denoted by $\phi(m_\star)$ and $q_{\rm Z}(m_\star)$
is the metal yield of stars of mass $m_\star$. From Fig.~1 it
is apparent that metallicity rapidly (in a time less than a few
$10^8$ Gyr) increases from primordial content to the limiting
value
\begin{equation}
Z_{\mathrm{cold}}\approx \frac {s}{s \gamma -1} A_{\rm Z},
\end{equation}
with $A_{\rm Z}=0.043$ for the adopted IMF ($A_{\rm Z}=0.021$ for the
Salpeter IMF).

In early galaxy evolution an important role is played by dust,
that absorbs the UV emission and re-radiates it in the mid and
far-IR band. The amount of dust in a galaxy is expected to be
correlated with that of cold gas [or SFR, see Eq.~(A3)] and
with metallicity. Mao et al. (2007) have shown that it is
possible to describe the luminosity-reddening relation found by
Shapley et al. (2001) for $z\approx 3$ LBGs with the simple law
\begin{equation}
A_{1350}\approx 0.35\,\left(\frac {\dot{M}_{\star}}{M_{\odot}\,
\mathrm{yr}^{-1}}\right)^{0.45}\,
\left(\frac{Z}{Z_{\odot}}\right)^{0.8},
\end{equation}
where $A_{1350}$ is the attenuation at $1350$ {\AA}. By
inserting in the above equation the SFR and metallicity, we get
the attenuation as function of time. Then the UV magnitude is
simply given by
\begin{equation}
M_{1350}\approx -18.6-2.5\,\log{\left({\dot{M}_{\star}\over M_{\odot}~\mathrm{yr}^{-1}}\right)} + A_{1350}~.
\end{equation}

This simplified treatment of dust attenuation proved to be quite a
good approximation for low-luminosity LBGs and LAEs, which exhibit
low attenuation (Mao et al. 2007).

On the other hand, for massive halos attenuation is large soon
after $10^8$ yr; this corresponds to the quick appearance of
very luminous submm-selected galaxies. For these systems the
model by Granato et al. (2004) includes a sophisticated
treatment of dust attenuation through the GRASIL code (Silva et
al. 1998).

The scheme presented here traces the evolution of single galaxy
halos as function of time, given their mass and their formation
redshift. The results can be interfaced to the formation rate,
$\mathrm{d}^2\, N_{\mathrm{ST}}/\mathrm{d} M_{\rm H}\,
\mathrm{d} t_{\rm z}$ (see \S~2.2) yielding LFs, counts and
redshift distribution for spheroidal galaxies. Moreover, the
quasar LF can also be reproduced (Lapi et al. 2006), as well as
the LF of LBGs and Ly$\alpha$ emitters (Mao et al. 2007). A
summary of the parameters used by the model and of its
achievements is presented in Tables 1 and 2 of Lapi et al.
(2006).
\end{appendix}

\end{document}